\documentclass[pop,aps,floats,preprint,graphics,amsmath,amssymb,superscriptaddress,showpacs,pre]{revtex4-1}
\usepackage{graphics,graphicx,color}
\begin{document}

\title{Scrape Off Layer profiles interpreted with filament dynamics}

\author{F. Militello} 
\affiliation{CCFE, Culham Science Centre, Abingdon, Oxon, OX14 3DB, UK}
\author{J.T. Omotani} 
\affiliation{CCFE, Culham Science Centre, Abingdon, Oxon, OX14 3DB, UK}

\begin{abstract}
A theoretical framework is developed to link the density profiles in the Scrape Off Layer (SOL) with the fluctuations (filaments) that generate them. The framework is based on the dynamics of independent filaments and their statistical behaviour and can be used to rigorously understand the mechanisms that lead to flattening and broadening of the SOL profiles as well as the radial increase of the relative fluctuation amplitude.  
\end{abstract}

\maketitle

\section{Introduction} \label{sec0}

The exhaust of power and particles in experimental magnetic fusion devices determines the level of interaction between the plasma and the material surfaces \cite{Loarte2007}. 

Experiments showed that, universally across machines, the midplane profiles of the density and of the temperature tend to flatten at a certain distance from the separatrix and close to the first wall \cite{Asakura1997,LaBombard2001,Lipschultz2002,Whyte2005,Lipschultz2005,Garcia2007,Carralero2014,Militello2016}. This leads to a stronger plasma surface interaction with the first wall rather than with the divertor components, which are specifically designed to sustain the large fluxes associated with the exhaust. This non-exponential nature of the profiles, which we call \textit{flattening} in the rest of the paper, led to the distinction between a near SOL, close to the separatrix, where the gradients are steep, and a far SOL, with slowly varying profiles and further out towards the wall \cite{LaBombard2001}. Another feature that appears to be universal to all the measurements is the response of the density profiles to increasing fuelling levels. Both in the near and far SOL, the decay length becomes longer at a higher fuelling level, but in the latter the change is much stronger, so that the two regions respond in a different way to the main plasma conditions. 

\begin{figure}
\includegraphics[height=6cm,width=6cm, angle=0]{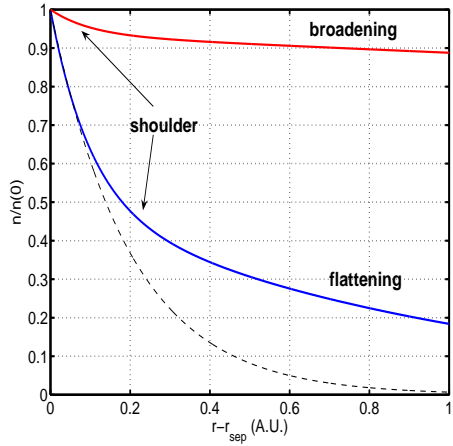}
\caption{Schematic representation of typical SOL profiles at low (blue curve) and high (red curve) fuelling. The dashed line represents a pure exponential decay and its used to show the effect of the flattening. The far SOL extends outwards from the shoulder position.}
\label{fig1}
\end{figure}

At the same time, as the fuelling level increases, the boundary between near and far SOL, called the \textit{shoulder}, moves closer to the separatrix \cite{LaBombard2001,Militello2016} and the near SOL shrinks accordingly. At high fuelling levels, the far SOL is almost flat and pervades most of the open field line region (if not all). We call this regime density \textit{broadening} in order to avoid confusion with the density \textit{flattening} of the far SOL, see Fig.\ref{fig1}.

Several experiments reported that the density profiles are also affected by the plasma current \cite{McCormick1992,Rudakov2005,Garcia2007b,Militello2016}. Higher currents correspond to steeper gradients globally and can even prevent the broadening at comparable line averaged density \cite{Militello2016}. It is possible that this effect is related to a reduction of the connection length, as the profiles significantly steepen in the wall shadow \cite{Lipschultz2002,Rudakov2005,Garcia2007}, where the field lines impinge on limiters rather than the divertor, with a step change reduction in the connection length. These results suggest that parallel physics plays a crucial role in determining the shape of the profiles. 

Interestingly, the electron temperature profiles show a certain degree of flattening but they typically do not respond to fuelling scans \cite{LaBombard2001,Lipschultz2002,Whyte2005}. This suggests that the electron temperature is not a crucial element for the density broadening and also that the temperature (heat) and density exhaust might be regulated by different mechanisms \cite{Militello2016}. A further consequence of this observation is that increased ionisation is generally unlikely to generate and sustain the broadened profiles as, if this was the case, the electron temperature should decrease. Indeed, the additional ionisation would remove energy from the colliding electrons, thus cooling them down for identical power crossing the separatrix. 

Experimental evidence shows a very turbulent SOL, with coherent filamentary structures erupting from the main plasma and transporting the plasma outwards. Visual cameras \cite{Kirk2006,Dudson2008,BenAyed2009}, gas puff imaging \cite{Zweben2002,Terry2003,Myra2006,Garcia2013} and upstream Langmuir probes \cite{LaBombard2001,Boedo2001,Garcia2007, Garcia2007b,Militello2013} show the presence of large fluctuations of the thermodynamic quantities in the SOL, especially far from the separatrix. It is therefore the time average over these structures that generates the SOL profiles, which are not equilibria in the proper sense. 

A lot of work was carried out to characterise these fluctuations, both from an experimental (see \cite{DIppolito2011} for a review) and a theoretical point of view \cite{Krasheninnikov2001,Garcia2004,Myra2006b,Militello2012,Militello2013b,Easy2014,Omotani2015,Easy2014,Militello2013,Militello2016,Ricci2014,Tamain2014}. Of particular interest is the observation that the turbulence seems to have a universal behaviour in the SOL, which can be captured by describing the statistics of the fluctuations through a gamma distribution \cite{Graves2005}, a result explained also by a recent theoretical model \cite{Garcia2012}. Also, results in literature suggest that the non Gaussian behaviour of the fluctuations is more evident in the far SOL than in the near SOL when profiles are flattened, and throughout the whole SOL when they are broadened \cite{Graves2005,Militello2013}. In addition, the relative amplitude of the fluctuations tends to increase with the radial position \cite{Terry2003,Zweben2015}, although this increase can be barely visible in certain instances \cite{Garcia2007}. Using gas puff imaging, both the waiting times of the filaments (i.e. the inverse of the average generation rate) and their amplitude were shown to be exponentially distributed \cite{Garcia2013}. Finally, a single filament generates a midplane Langmuir probe signal that has a characteristic double exponential shape \cite{Boedo2001,Antar2003,Militello2013}, which can be traced to its radial variation \cite{Garcia2006}. 

In this paper, we present a theoretical framework that elucidates how filament dynamics generate SOL profiles, thus unifying the two fields of research. The crucial insight obtained from our work is that filamentary dynamics can describe both the near and far SOL behaviour in a consistent way, both in terms of density profiles and fluctuation statistics. Also, the framework can allow first principles predictions for the SOL response to changes in the plasma conditions, if filament dynamics is properly constrained with experimental data or theoretical models. Our work can be seen as an extension to the radial dimension of the statistical treatment in \cite{Garcia2012} and therefore inherits all its predictive capabilities. A similar pioneering attempt to relate fluctuations to profiles is described in \cite{DIppolito2002}, where a reduced version of our approach was presented (only large blobs in the far SOL were examined and with a less sophisticated statistical approach). 

\section{Theoretical Framework} \label{sec1}

We start with the assumption that in the radial direction, $x$, individual filaments have a well defined shape, $\Lambda(x,w)$, defined by only one parameter, $w$, which measures the perpendicular width of the filament. We set the separatrix position at $x=0$ and define positive values of $x$ to represent the SOL region. Now, the filaments are subject to radial motion \cite{Krasheninnikov2001,Garcia2006,Easy2014} and to draining due to parallel losses, which can be represented in the Eulerian frame as:
\begin{equation}
\label{2} 
\eta_i(x,t)=\eta_{0,i}F_i(t)\Lambda\left(x-\int_0^t V_i(t')dt',w_i\right) 
\end{equation}
where $t$ is time, $\eta_0$ represents the initial amplitude of the thermodynamical variable associated with the filament (i.e. density or temperature), $F(t)$ describes the reduction of this amplitude due to generic parallel losses, $V(t)$ is the radial velocity of the filament and the subscript $i$ is a label representing the fact that different filaments can have different properties. Note that we assume that both $F$ and $V$ have their $i$ dependence only through the initial amplitude and size of the filament, i.e. $V_i(t)=V(t,\eta_{0,i},w_i)$. The form of Eq.\ref{2} guarantees invariance of the filament shape under radial translations (i.e. $x-\int_0^t V(t') dt'=const$), which means that we implicitly assume that the shape is rescaled self similarly. Also, Eq.\ref{2} represents a filament that crosses the sepatratrix (i.e. $x=0$) at $t=0$. 

In an ideal statistical steady state, filaments are continuously emitted from the separatrix. We further assume that they do not interact with each other while travelling in the SOL even if new filaments move in the wake of older filaments so we simply add their amplitude to generate the time signals. In other words, at each time $t$ and position $x$ the train of filaments would produce a signal given by:
\begin{equation}
\label{3}
\theta(x,t) = \sum_{i=1}^\infty \eta_i\left(x,t-t_i\right),
\end{equation}
where $t_i$ is the time at which the ith filament crosses the separatrix. At each radial position, $x$, the signal produced by Eq.\ref{3} is also known as shot noise and it has the features of a Poisson process \cite{Pecseli2000,Garcia2012} if the $t_i$ are uniformly distributed. 

The final step to go from the filaments' motion to the radial profile of the thermodynamic variable, $\Theta(x)$, is to time average $\theta(x,t)$:
\begin{equation}
\label{4}
\Theta(x) =   \overline{\sum_{i=1}^{\infty}\eta_i(x,t-t_i)}
\end{equation}
where we have defined $\overline{\cdots}\equiv \displaystyle{\lim_{\Delta T \rightarrow \infty}}\Delta T^{-1}\int_{0}^{\Delta T}\cdots dt$. 

The problem now is that we need to reach statistically meaningful conclusions regarding the profile behaviour. In other words, if we knew exactly the values of all the quantities with the $i$ subscript for all the filaments, calculating Eq.\ref{4} would be trivial, but since we only have their statistical distributions, also the profile should be interpreted as a statistical quantity. In this respect, it is useful to start by calculating $\Theta(x)$ in a finite interval $\Delta T$, in which $K$ filaments (and not infinity) will contribute to the signal. The number of filaments is Poisson distributed due to the assumptions of independence between the filaments and of constant generation rate. We therefore have that:
\begin{equation}
\label{4aa}
\Theta_{\Delta T}(x) = \frac{1}{\Delta T}\int_0^{\Delta T}dt \sum_{i=1}^K \eta_i(x,t-t_i)
\end{equation}  
In general, this is an ill defined quantity, since $K$ is a statistical variable that depends on $\Delta T$. Since the process described is ergodic by construction \cite{Pecseli2000}, the time average corresponds to an ensemble average over the possible statistical outcomes. This leads to:
\begin{equation}
\label{4bb}
\Theta_{\Delta T}(x) = <\theta(x,t)>=\int_0^{\infty}d\eta_{0,i}P_{\eta_0}\int_0^{\infty}dw_iP_w \sum_{K=1}^\infty P_K\sum_{i=1}^K \int_0^{\Delta T} dt_i P_{t_i}\eta_i(x,t-t_i),
\end{equation}
where $P_{\eta_0}$ and $P_w$ represent the probability distribution functions of the initial amplitudes and of the widths, $P_K=\lambda^K e^{-\lambda}/K!$, $\lambda=\Delta T/\tau_w$ and $\tau_w$ is the average time between filaments, i.e their waiting time (inverse of the emission rate or rate of the Poisson process) and $P_{t_i}=1/\Delta T$ is the homogeneous probability distribution of the arrival times associated with a Poisson process. The operator $<\cdots>$ represents an ensemble average. It can be shown \cite{Pecseli2000} that for the process of interest, $\sum_{K=1}^\infty P_K\sum_{i=1}^K\int_0^{\Delta T} dt_i P_{t_i}\eta_i(x,t-t_i)=\tau_w^{-1}\int_{-\infty}^{\infty}dt\eta_i(t)$ when $\Delta T\rightarrow \infty$.

We can then write:
\begin{equation}
\label{4b}
\Theta(x) = \{\eta(x,t)\}=\frac{1}{\tau_w} \int_{-\infty}^{\infty}dt\int_0^{\infty}d\eta_0 \int_0^{\infty}dw \left[\eta(x,t) P_{\eta_0}(\eta_0)P_w(w)\right],
\end{equation}
where we dropped the subscripts and used curly brackets to define the average operator. Note that the order in which the integrals are performed is not relevant since $\eta_0$ and $w$ do not depend on time or on each other (the latter statement is an assumption). 


To compare with experimental observations, it is useful to determine other properties of the signals, such as successive statistical moments. These, of course, characterise more accurately the time series and provide a more stringent constraint for model validation. After the time average given in Eq.\ref{4b}, it is natural to study the variance of the signal, which broadly speaking represents the amplitude of the fluctuations. This is defined as:
\begin{equation}
\label{4c}
\sigma(x) = \overline{\left[\sum_{i=1}^{\infty} \eta_i(x,t-t_i)-\Theta(x)\right]^2}=\overline{\left[\sum_{i=1}^{\infty} \eta_i(x,t-t_i)\right]^2}-\Theta(x)^2.
\end{equation}
Applying the same statistical procedure as above to Eq.\ref{4c}, we obtain: $\sigma(x) =\{\eta(x,t)^2\}$. A proof of this is given in \cite{Pecseli2000}, together with similar expressions for higher order moments like the skewness: $\{\eta(x,t)^3\}/\{\eta(x,t)^2\}^{3/2}$, and the kurtosis: $\{\eta(x,t)^4\}/\{\eta(x,t)^2\}^{2}$.

\section{Application of the framework}

We describe here the simplest application of the framework, for filaments with a single amplitude $\eta_*$ and width $w_*$, corresponding to $P_{\eta_0}(\eta_0)=\delta(\eta_0-\eta_*)$ and $P_w(w)=\delta(w-w_*)$ where $\delta$ is the Dirac delta function. Taking a single timescale for the parallel draining, $F(t)=e^{-t/\tau}$, and a simple shape function with a vertical leading edge and exponential tail, $\Lambda(x,w)=e^{x/w}H(-x)$, where $H(x)$ is the Heaviside step function, we have:
\begin{equation}
\label{14}
\Theta(x) = \frac{\eta_*e^{\frac{x}{w_*}}}{\tau_w} \int_{-\infty}^{\infty}dt e^{-\frac{t}{\tau}-\frac{X(t)}{w_*}}H[X(t)-x]=\frac{\eta_*e^{\frac{x}{w_*}}}{\tau_w} \int_{X^{-1}(x)}^{\infty}dt e^{-\frac{t}{\tau}-\frac{X(t)}{w_*}}.
\end{equation}
With constant filament velocity, $X(t)= \int_0^t V(t')dt'= Vt$, it is easy to prove that:
\begin{equation}
\label{14b}
\Theta(x) = \frac{\eta_*}{\frac{\tau_w}{\tau}\left(1+\frac{V\tau}{w_*}\right)}e^{-\frac{x}{V\tau}},
\end{equation} 
and:
\begin{equation}
\label{14c}
\frac{\sigma(x)^{1/2}}{\Theta(x)} = \frac{\sqrt{2}}{2}\sqrt{\frac{\tau_w}{\tau}\left(1+\frac{V\tau}{w_*}\right)}.
\end{equation} 

Note that Eq.\ref{14b} would remain the same also under the assumption that $P_{\eta_0}(\eta_0)=\eta_*^{-1}e^{-\eta_0/\eta_*}$, while Eq.\ref{14c} would not have the $\sqrt{2}/2$ factor on the right hand side. The presence of filaments of different sizes can be taken into account by multiplying $\Theta(x)$ and $\sigma(x)$ by a suitable probability distribution function and integrating over all the possible realisations. If, for example, we take a log-normal distribution, the results obtained above remain basically identical if we replace $w_*$ with the average value of the distribution. 

The constant velocity model captures the naive approach used in several empirical interpretations of the profile broadening: a transition in the filament dynamics that produces larger perpendicular velocities leads to broader profiles. Our analysis shows that this approach cannot explain relevant experimental observations associated with the flattening of the profiles (the decay length is constant) and with the radially changing statistics (which remain the same). 

Despite its limitations, the constant velocity model is presented here to describe how the framework can be used and because it can help to identify the mechanisms leading to the flattening and broadening (see next Section). We also developed advanced models with more realistic assumptions for $F$, $V$ and the probability density functions which will be presented in a separate publication. They are not shown here because they are outside the scope of this Letter, but they are at the base of the discussion in the next Section.

\section{Discussion}

If the constant velocity model is interpreted in a local sense, it allows the parametric dependences that are needed to induce flattening and broadening in the density profile to be identified (we do not discuss here the temperature profiles, although the model could capture their behaviour as well). In particular, it is easy to see that the profile decay length depends on the combination $V\tau$, which means that at least one of these quantities must change in time if the flattening is to be recovered (note that there is no direct filament size dependence in this expression). 

This can be easily achieved if the filaments accelerate in the SOL, as an increasing velocity translates to a longer decay length, but also implies larger relative amplitudes of the perturbations in the far SOL, see Eq.\ref{14c}, consistent with the experimental observations. The experimental literature is often contradictory when it comes to the radial variation of the filament's velocity. Increasing \cite{Zweben2015,Zweben2016}, decreasing \cite{Myra2006} and unchanging \cite{Kirk2016,Militello2016} velocities were reported, although recent results based on a large database of filaments in NSTX suggest that the filaments indeed accelerate \cite{Zweben2016}. From a theoretical perspective, a filament might accelerate because of the reduced background through which it moves \cite{Omotani2015} or because of the increased resistivity which reduces the sheath dissipation \cite{Easy2016}. Albeit the simplest interpretation, an increasing velocity is a sufficient but not necessary condition for the profile flattening. 

An alternative mechanism is the fact that the upstream draining of the filament's density can occur on multiple time scales, if we consider the fast relaxation of the parallel gradients arising from the ballooning nature of the initial perturbation and the slower exhaust at the target. The former time scale is proportional to the length scale of the parallel variations in the filament, and the latter to the midplane to target connection length (they can be separated by a factor $5-10$). While this interpretation can account for the flattening of the profile, it is incompatible with the experimental observation that the relative amplitude of the fluctuations increases moving outwards. Indeed, Eq.\ref{14c} reveals that $\tau$ increasing away from the separatrix implies a decreasing $\sigma^{1/2}/\Theta$. On the other hand, it would be sufficient to assume that the ratio $\tau_w/\tau$ remains constant (or increases) to counteract this effect and retrieve again a growing fluctuation amplitude. This could be easily achieved if some of the filaments did not reach the far SOL, which could be the case for small amplitude and/or size perturbations which could dissipate in the neighbourhood of the separatrix, thus leaving gaps in the sequence of filaments.     

Most likely, the cause of the flattening is a combination of the above mentioned effects, but the framework developed in our work allows discussing rigorously how these effects combine. Similarly, the occurrence of the broadening of the profile would require more intense acceleration of the filaments which would lead to higher velocities. However, in order not to lower separatrix density, a simultaneus increase in the filament amplitude would be required. These are directly testable consequences of the model. On the other hand, it is also possible that the plasma exhaust can become less efficient as the line average density increases (i.e. $\tau$ increases in the broadened region). This could well be caused by charge exchange interactions between neutral particles and ions, which can `clog' the filament motion towards the target. As line average density increases in the core, so does the neutral density in the SOL \cite{Militello2016}, which leads to a greater probability of collisions. In contrast to ionisation, this plasma-neutrals interaction does not cool down the electrons and is therefore consistent with the observed unchanging electron density profiles. Most of the interaction is expected in the divertor region, where the neutral density is higher. 

\section{Summary and Conclusions}

We presented a statistical framework that rigorously describes how filaments form average profiles in the SOL and determine higher order moments. The inputs of the framework are experimentally or theoretically derived models for the draining towards the divertor and radial motion of single filaments, as well as the statistics of their initial amplitudes and radial width. Here, we applied the framework to the simplest case of exponential draining and constant velocity, which enables us to identify the parameter combinations that control the behaviour of the statistical moments. On the other hand, it is clear that more sophisticated input models are required to fully understand the physics of the density flattening and broadening in experiments. The development of this framework motivates further experimental and theoretical studies aimed at identifying the dependences on the main plasma parameters of $F$, $V$ and the probability density functions for amplitudes and widths of the filaments. If such dependences were clearly identified, the framework could be used to obtain reliable predictions for next generation machines.     

\acknowledgements

F.M. acknowledges useful discussions with Dr. Andrew Kirk, Dr. N. Walkden, Mr. T. Farley and Mr. L. Easy. We also thank Dr. Y. Liu for carefully reading the manuscript. This work was funded by the RCUK Energy Programme [grant number EP/I501045]. To obtain further information on the data and models underlying this paper, whose release may be subject to commercial restrictions, please contact PublicationsManager@ccfe.ac.uk .

\end{document}